\documentclass[prl,twocolumn,showpacs]{revtex4}
\usepackage{epsfig}
\usepackage{amsbsy}
\usepackage{amsmath}
\usepackage{amsfonts}
\usepackage{graphicx}

\begin{document}

\newcommand{\bqa}{\begin{eqnarray}}
\newcommand{\eqa}{\end{eqnarray}}
\newcommand{\nn}{\nonumber}
\newcommand{\nl}[1]{\nn \\ && {#1}\,}
\newcommand{\erf}[1]{Eq.\ (\ref{#1})}
\newcommand{\dg}{^\dagger}
\newcommand{\ip}[1]{\left\langle{#1}\right\rangle}
\newcommand{\bra}[1]{\langle{#1}|}
\newcommand{\ket}[1]{|{#1}\rangle}
\newcommand{\braket}[2]{\langle{#1}|{#2}\rangle}
\newcommand{\st}[1]{\langle {#1} \rangle}
\newcommand{\etal}{{\it et al.}}
\newcommand{\half}{{\small \frac 12}}

\title{Optimal Remote State Preparation}
\author{Dominic W.\ Berry and Barry C.\ Sanders}
\affiliation{Department of Physics and
        Centre for Advanced Computing -- Algorithms and Cryptography,   \\
        Macquarie University,
        Sydney, New South Wales 2109, Australia}
\date{\today}

\begin{abstract}
We prove that it is possible to remotely prepare an ensemble of
non-commuting mixed states using communication equal to the Holevo information
for this ensemble. This remote preparation scheme may be used to convert between
different ensembles of mixed states in an asymptotically lossless way, analogous
to concentration and dilution for entanglement.
\end{abstract}
\maketitle

In classical information theory one of the central problems is that of coding.
From Shannon's noiseless coding theorem \cite{shannon}, if message $i$ is given
with probability $p_i$, then a sequence of messages may be compressed to an
average number of bits per message equal to the Shannon entropy $H=\sum_i p_i
\log_2 p_i$. This result means that classical communication of
$H$ bits per message is sufficient to reconstruct the sequence of messages;
conversely $H$ bits of communication per message may be obtained. Thus the
Shannon entropy may be given a definite interpretation as the classical
information per message. Here we show that, in the quantum case where density
$\rho_i$ is given with probability $p_i$, it is possible to give a similar
interpretation to the Holevo information. The Holevo information is given by
$S( \sum_i p_i \rho_i )-\sum_i p_i S(\rho_i)$, where $S(\rho)=-{\rm Tr}(\rho
\log_2\rho)$ is the von Neumann entropy. It has previously been shown
\cite{kholevo,schuwest,holevo} that it is possible to perform classical
communication equal to the Holevo information.  We show that it is possible to
effectively reverse this process, and remotely prepare these states using
communication equal to the Holevo information. This result means that it is
possible to convert from ensembles of mixed states to classical information and
back again in an asymptotically lossless way, analogous to concentration and
dilution of entanglement \cite{condil}.

In remote state preparation \cite{lo00,pati,bennett,devetak,leung}, Alice (A)
wishes to prepare state $\rho_i$ with probability $p_i$ in the laboratory of Bob
(B). Because this ensemble of mixed states ${\sf E}=\{p_i,\rho_i\}$ may be used
to perform communication equal to its Holevo information \cite{schuwest,holevo},
the Holevo information provides a lower bound to the communication required for
remote state preparation \cite{bennett}. The problem of remote
state preparation at this lower bound has hitherto only been solved for the
special case that all densities to be prepared commute \cite{dur}.

It is instructive to first summarize a non-optimal scheme for preparing a single
state from Ref.\ \cite{bennett}. In order to approach this problem, it is
convenient to consider preparation of a pure state $\ket{\Phi_i}$
shared between Alice and Bob, such that Bob's reduced density matrix is
$\rho_i$. We denote this by the ensemble ${\cal E}=\{p_i,\ket{\Phi_i}\}$. The
state $\ket{\Phi_i}$ has Schmidt decomposition
\begin{equation}
\ket{\Phi_i}=\sum_{j\in D} \sqrt{\lambda_i^j}\ket{\varphi_i^j}_A
\ket{\chi_i^j}_B.
\end{equation}
Both modes are of dimension $d$, and $D=\{1,\cdots,d\}$.
Preparing this state is equivalent to remotely preparing the mixed state
$\rho_i=\sum_{j\in D} \lambda_i^j \ket{\chi_i^j}_B\bra{\chi_i^j}$.

In order to remotely prepare $\rho_i$, Alice consumes entanglement and
performs classical communication to Bob. Via local operations on Alice's side,
any maximally entangled state may be brought to the form
\begin{equation}
\ket{\Theta_i}=\frac 1{\sqrt d}\sum_{j\in D}
\ket{\varphi_i^j}_A\ket{\chi_i^j}_B.
\end{equation}
Alice then performs a measurement described by a positive operator-valued
measure (POVM) with two elements, $\Pi_i^1=\frac 1{\Lambda_i} \sum_{j\in D}
\lambda_i^j \ket{\varphi_i^j}_A\bra{\varphi_i^j}$ and $\Pi_i^0=\openone-
\Pi_i^1$, where $\Lambda_i=\max_j\{\lambda_i^j\}$. If the
measurement result is 1, the resulting (unnormalized) state is
\begin{equation}
\big(\sqrt{\Pi_i^1}\otimes \openone\big)\ket{\Theta_i}=
\frac 1{\sqrt{\Lambda_i d}}\sum_{j\in D}
\sqrt{\lambda_i^j}\ket{\varphi_i^j}_A\ket{\chi_i^j}_B.
\end{equation}
Normalization gives the state Alice wished to prepare, $\ket{\Phi_i}$.
This result occurs with probability $1/(\Lambda_i d)$. The state resulting from
a measurement result of 0 is not usable, and this measurement result counts as a
failure.

The preparation of the state therefore requires the classical communication of
the number of the measurement result that is a success. Let us assume that the
measurement is performed a maximum of $M$ times, where $M$ is the smallest
integer not less than $\Lambda d\ln(\Lambda d)$, for
$\Lambda=\max_i\{\Lambda_i\}$. If there is a success, then the number of the
success is communicated; otherwise zero is communicated. As there are no more
than $M+1$ alternative messages to communicate, the number of bits required is
$\log(\Lambda d)+O[\log\log(\Lambda d)]$. Throughout this letter we denote
logarithms to base 2 by log and logarithms to base $e$ by ln.

The probability of all the measurements being failures is
$\big(1-1/\Lambda_i d\big)^M\le 1/\Lambda d$.
If there is a success the fidelity is equal to 1, so the average fidelity must
be at least $1-1/\Lambda d$. Therefore, in the limit of large $\Lambda d$, the
communication required is $\log(\Lambda d)+O[\log\log(\Lambda d)]$, and the
average fidelity is arbitrarily close to 1.

The state preparation protocol of Ref.\ \cite{bennett} does not, in general,
reach the Holevo limit. In order to reach the Holevo limit, we generalize this
state preparation scheme to jointly prepare a number of states. That is, we
prepare the tensor product of $n$ states
$\ket{\Phi_u}=\ket{\Phi_{i_1}}\otimes\cdots\otimes\ket{\Phi_{i_n}}$
with probability $p_u=p_{i_1}\times\cdots\times p_{i_n}$. Here we use the
notation $u=(i_1,\cdots,i_n)$. This tensor product state has the Schmidt
decomposition
\begin{equation}
\label{prepthis}
\ket{\Phi_u} = \sum_{J\in D^n} \sqrt{\lambda_u^J}
\ket{\varphi_u^J}_A\ket{\chi_u^J}_B,
\end{equation}
where
$\lambda_u^J =\lambda_{i_1}^{j_1}\times \cdots \times \lambda_{i_n}^{j_n}$,
$\ket{\varphi_u^J}_A = \ket{\varphi_{i_1}^{j_1}}_A \otimes \cdots \otimes
\ket{\varphi_{i_n}^{j_n}}_A$, and
$\ket{\chi_u^J}_B = \ket{\chi_{i_1}^{j_1}}_B \otimes \cdots \otimes
\ket{\chi_{i_n}^{j_n}}_B$.

We introduce the subspace of $\ket{\Phi_u}$
\begin{equation}
B_u = \{ J: \lambda_u^J < 2^{-n(\bar S-\delta)} \},
\end{equation}
where $\bar S = \sum_i p_i S(\rho_i)$. We will also use the notation
$\bar\rho=\sum_i p_i \rho_i$, so the Holevo information for the
ensemble is $S(\bar\rho)-\bar S$.
Given any positive $\epsilon$ and $\delta$, there exists $n_2(\epsilon,\delta)$
such that, for all $n>n_2(\epsilon,\delta)$,
\begin{equation}
\label{fidel}
\left\langle \sum_{J\in B_u} \lambda_u^J \right\rangle \ge 1-\epsilon,
\end{equation}
where the expectation value indicates the average over $u$ with probabilities
$p_u$. In Ref.\ \cite{holevo}, this result is given for the subspace
$B'_u = \{ J: 2^{-n(\bar S+\delta)} < \lambda_u^J < 2^{-n(\bar S-\delta)} \}$.
The subspace we use includes additional small values of $\lambda_u^J$, which
can only increase the sum. Therefore the result (\ref{fidel})
must hold for the subspace $B_u$.

Now we introduce a POVM with elements
\begin{equation}
\Pi_u^1=2^{n(\bar S-\delta)}\sum_{J\in B_u} \lambda_u^J
\ket{\varphi_u^J}_A\bra{\varphi_u^J}
\end{equation}
and $\Pi_u^0=\openone-\Pi_u^1$. As above, a maximally entangled state shared
between Alice and Bob may be brought, via local operations on Alice's side, to
the form
\begin{equation}
\ket{\Theta_u}= \frac 1{d^{n/2}}\sum_{J\in D^n}
\ket{\varphi_u^J}_A\ket{\chi_u^J}_B.
\end{equation}
Alice performs a measurement described by the above POVM on this entangled
state. After a measurement which yields the result 1, the resulting unnormalized
state is
\begin{equation}
(\sqrt{\Pi_u^1}\otimes \openone)\ket{\Theta_u}=\frac{2^{n(\bar S-\delta)/2}}
{d^{n/2}} \sum_{J\in B_u} \sqrt{\lambda_u^J}\ket{\varphi_u^J}_A
\ket{\chi_u^J}_B.
\end{equation}
With normalization the state may be written as
\begin{equation}
\ket{\Phi'_u}=\frac 1{\cal N} \sum_{J\in B_u} \sqrt{\lambda_u^J}
\ket{\varphi_u^J}_A \ket{\chi_u^J}_B ,
\end{equation}
where ${\cal N}={\sqrt{\sum_{J\in B_u} \lambda_u^J}}$ is a normalization factor.
This state is not exactly equal to the state that was to be prepared
(\ref{prepthis}); however, the fidelity is
\begin{equation}
F_u =|\braket{\Phi_u}{\Phi'_u}|^2
= \frac 1{{\cal N}^2}\left| \sum_{J\in B_u} \lambda_u^J\right|^2
=\sum_{J\in B_u} \lambda_u^J.
\end{equation}
Using Eq.\ (\ref{fidel}), we find that $\langle F_u \rangle \ge 1-\epsilon$.

The dimension of the space used is $d^n$, and the maximum $\lambda_u^J$ is no
greater than $2^{-n(\bar S-\delta)}$. Therefore, from the above discussion for
the preparation of a single state, the state $\ket{\Phi_u'}$ can be prepared
with probability of success at least $1-d^{-n} 2^{n(\bar S-\delta)}$ and
communication $n(\log d-\bar S+\delta)+O(\log n)$. The average fidelity with the
state $\ket{\Phi_u}$ must therefore be at least
$1-\epsilon-d^{-n}2^{n(\bar S-\delta)}$.

The average fidelity of the reduced density matrices in Bob's mode must also be
at least $1-\epsilon-d^{-n}2^{n(\bar S-\delta)}$ due to the relation
\cite{jozsa} $F(\rho,\sigma) = \max_{\ket{\psi},\ket{\phi}}|\braket{\psi}
{\phi}|^2$, where $\rho$ and $\sigma$ are density matrices and $\ket{\psi}$ and
$\ket{\phi}$ are purifications of $\rho$ and $\sigma$.

We therefore see that the states $\ket{\Phi_u}$, or the corresponding reduced
density matrices for Bob, may be prepared with average fidelity arbitrarily
close to 1 and communication per prepared state arbitrarily close to
$\log d-\bar S$. This communication is still larger, in general, than the Holevo
bound of $S(\bar \rho)-\bar S$. In order to reach this bound we combine this
protocol with what is effectively Schumacher compression \cite{schu}.

Using Schumacher compression, the ensemble of states to be prepared, ${\sf E}$,
may be compressed to a space of dimension $2^{S(\bar \rho)}$. These states may
therefore be prepared in the above way with communication $S(\bar \rho)-\bar S$.
We will now show that this compression may be applied to the preparation of
entangled states, and with fidelity arbitrarily close to 1.

In order to apply Schumacher compression, we use a method similar to that
of Lo \cite{lo}. Lo shows that, given an ensemble of density matrices
${\sf E}=\{p_i,\rho_i\}$ to be transmitted, for any $\epsilon, \delta>0$, there
exists an $n$ such that the sequence of density matrices $\rho_u=\rho_{i_1}
\otimes\cdots\otimes\rho_{i_n}$ may be compressed to $S(\bar\rho)+\delta$ qubits
with average `distortion' less than $\epsilon$.

It is straightforward to modify Lo's derivation so that it deals with ensembles
of entangled states, and the fidelity is used rather than the distortion. For
simplicity we consider preparation of the state
\begin{equation}
\label{same}
\ket{\Phi_u} = \sum_{J\in D^n} \sqrt{\lambda_u^J}
\ket{\chi_u^J}_A\ket{\chi_u^J}_B.
\end{equation}
This state may be brought to the form (\ref{prepthis}) via unitary operations
on Alice's mode. As explained in Ref.\ \cite{holevo}, for all
$\epsilon,\delta>0$ there is an $n_1(\epsilon,\delta)$ such that, for all
$n>n_1(\epsilon,\delta)$, ${\rm Tr}(\bar\rho^{\otimes n} P)>1-\epsilon$,
where $P$ is a projector onto a space of dimension $2^{n[S(\bar\rho)+\delta]}$.
We will denote this space by $\Xi$, and write $\ket{\chi_u^J}$ in the form
\begin{equation}
\ket{\chi_u^J}=\alpha_u^J \ket{l_u^J} + \beta_u^J \ket{m_u^J},
\end{equation}
where $\alpha_u^J,\beta_u^J\ge 0$, $(\alpha_u^J)^2+(\beta_u^J)^2=1$, and
the states $\ket{l_u^J}$ and $\ket{m_u^J}$ are in the spaces $\Xi$ and
$\Xi^\perp$, respectively.

Now we replace each state (\ref{same}) with
\begin{equation}
\ket{\widetilde\Phi_u} = \sum_{J\in D^n} \sqrt{\lambda_u^J} (\alpha_u^J)^2
\ket{l_u^J}_A\ket{l_u^J}_B + {\cal B}_u \ket{l_u^0}_A\ket{l_u^0}_B .
\end{equation}
The state $\ket{l_u^0}$ may be chosen arbitrarily. The value of the coefficient
${\cal B}_u$ is chosen such that ${\cal B}_u\braket{\Phi_u}{l_u^0}_A
\ket{l_u^0}_B$ is a positive real number and normalization is preserved.
%The average fidelity is then given by $\sum_u p_u |\braket{\Phi_u}
%{\widetilde\Phi_u}|^2$.
We find that
\begin{align}
\braket{\Phi_u}{\widetilde\Phi_u}&=\sum_{J,J'}\sqrt{\lambda_u^J\lambda_u^{J'}}
(\alpha_u^J)^2(\alpha_u^{J'})^2 \braket{l_u^J}{l_u^{J'}}^2 \nn \\
&~~~+ {\cal B}_u\braket{\Phi_u}{l_u^0}_A\ket{l_u^0}_B \nn \\
&\ge \sum_J \lambda_u^J (\alpha_u^J)^4 \nn \\ &~~~
+\sum_{J\ne J'}\sqrt{\lambda_u^J \lambda_u^{J'}}(\alpha_u^J)^2
(\alpha_u^{J'})^2\braket{l_u^J}{l_u^{J'}}^2 \nn \\
&= \sum_J \lambda_u^J \left[ 1-2(\beta_u^J)^2+(\beta_u^J)^4 \right] \nn \\ & 
~~~+ \sum_{J\ne J'} \sqrt{\lambda_u^J\lambda_u^{J'}}(\beta_u^J)^2
(\beta_u^{J'})^2\braket{m_u^J}{m_u^{J'}}^2 \nn \\
&= 1-2\sum_J \lambda_u^J (\beta_u^J)^2 \nn \\ &~~~+
\sum_{J,J'} \sqrt{\lambda_u^J\lambda_u^{J'}}(\beta_u^J)^2(\beta_u^{J'})^2
\braket{m_u^J}{m_u^{J'}}^2 \nn \\
& \ge 1-2\sum_J \lambda_u^J (\beta_u^J)^2.
\end{align}
The average fidelity is therefore
\begin{align}
F& \ge \sum_u p_u \left|1-2\sum_J \lambda_u^J (\beta_u^J)^2\right|^2 \nn \\
&\ge \sum_u p_u \left[1-4\sum_J \lambda_u^J (\beta_u^J)^2\right]
\ge 1-4\epsilon.
\end{align}
In the last line we have used ${\rm Tr}(\bar\rho^{\otimes n}P)>1-\epsilon$.

\begin{table*}
\caption{Three analogous processes for classical communication vs entanglement. 
In boldface is shown the optimal remote state preparation discussed in this
letter, as well as the conversion between ensembles, discussed in
Ref.\ \cite{ben}, for which optimal remote state preparation is required.
\label{table}}
\begin{tabular}{|c|c|} \hline
Classical Communication & Entanglement \\ \hline
communication equal to the Holevo information & entanglement concentration \\ 
{\bf optimal remote state preparation } & entanglement dilution \\
{\bf conversion between ensembles} & conversion between entangled states \\
\hline
\end{tabular}
\end{table*}

In addition to $n>n_1(\epsilon,\delta)$, we take $n>n_2(\epsilon,\delta)$
and $n>1/\delta$, so the Schmidt coefficients for $\ket{\Phi_u}$
satisfy Eq.\ (\ref{fidel}). Now the modified states $\ket{\widetilde\Phi_u}$
have the Schmidt decompositions
$\ket{\widetilde\Phi_u} = \sum_{J\in D^n}\sqrt{\widetilde\lambda_u^J}
\ket{\widetilde\chi_u^J}_A\ket{\widetilde\chi_u^J}_B$.
We introduce the subspace
\begin{equation}
\widetilde B_u = \{ J: \widetilde\lambda_u^J < 2^{-n(\bar S-2\delta)} \} .
\end{equation}
Now we may place limits on the sum over the $\widetilde\lambda_u^J$ in the
following way:
\begin{align}
&\sum_{J\in \widetilde B_u}\widetilde\lambda_u^J =
1-\!\! \sum_{J\in \widetilde B_u^{\perp}}\widetilde\lambda_u^J \nn \\
&= 1-\!\!\sum_{J\in \widetilde B_u^{\perp}} \bra{\widetilde\chi_u^J} \!
\left[\widetilde \lambda_u^J\ket{\widetilde\chi_u^J}\bra{\widetilde\chi_u^J} -
\!\! \sum_{J'\in D^n}\lambda_u^{J'}\ket{\chi_u^{J'}}\bra{\chi_u^{J'}}
\right] \! \ket{\widetilde\chi_u^J} \nn \\ & ~~~ -\!\!\sum_{J\in \widetilde
B_u^{\perp}} \sum_{J'\in D^n}\lambda_u^{J'} |\braket{\chi_u^{J'}}
{\widetilde\chi_u^J}|^2 \nn \\
&\label{above} = 1- {\rm Tr}[\widetilde P(\widetilde\rho_u-\rho_u)]-\!\!
\sum_{J\in\widetilde B_u^{\perp}}\!\sum_{J'\in D^n}
\lambda_u^{J'} |\braket{\chi_u^{J'}}{\widetilde\chi_u^J}|^2.
\end{align}
Here $\widetilde P$ is the projector $\sum_{J\in \widetilde B_u^{\perp}}
\ket{\widetilde\chi_u^J}\bra{\widetilde\chi_u^J}$, $\widetilde\rho_u$ is
Bob's reduced density operator for state $\ket{\widetilde\Phi_u}$ and
$\widetilde B_u^\perp=D^n\backslash \widetilde B_u$. Using results
from Ref.\ \cite{nielsen} we have the inequalities
\begin{align}
{\rm Tr}[\widetilde P(\widetilde\rho_u-\rho_u)]&\le D(\widetilde\rho_u,\rho_u)
\nn \\\le\sqrt{1-F(\widetilde\rho_u,\rho_u)} &\le\sqrt{1-F(\ket{\widetilde
\Phi_u},\ket{\Phi_u})},
\end{align}
where $D(\widetilde\rho_u,\rho_u)=\frac 12 {\rm Tr}|\widetilde\rho_u-\rho_u|$
is the trace distance. Note that the fidelity defined in Ref.\ \cite{nielsen} is
the square root of the fidelity defined here. The third term on the right hand
side of Eq.\ (\ref{above}) may be evaluated as
\begin{align}
& \sum_{J\in\widetilde B_u^{\perp}}\sum_{J'\in D^n}
\lambda_u^{J'} |\braket{\chi_u^{J'}}{\widetilde\chi_u^J}|^2 \nn \\ &=
\sum_{J\in\widetilde B_u^{\perp}}\sum_{J'\in B_u}
\lambda_u^{J'} |\braket{\chi_u^{J'}}{\widetilde\chi_u^J}|^2 + \!
\sum_{J'\in B_u^{\perp}} \! \lambda_u^{J'} \!\! \sum_{J\in\widetilde B_u^{\perp}}
|\braket{\chi_u^{J'}}{\widetilde\chi_u^J}|^2 \nn \\ &\le
\sum_{J\in\widetilde B_u^{\perp}}2^{-n(\bar S-\delta)}\sum_{J'\in B_u}
|\braket{\chi_u^{J'}}{\widetilde\chi_u^J}|^2 +
\sum_{J'\in B_u^{\perp}}\lambda_u^{J'} \nn \\
&\le \sum_{J\in\widetilde B_u^{\perp}}2^{-n(\bar S-\delta)} +
\sum_{J\in B_u^{\perp}}\lambda_u^{J} \nn \\
&= 2^{-n\delta}\sum_{J\in\widetilde B_u^{\perp}}2^{-n(\bar S-2\delta)} +
\sum_{J\in B_u^{\perp}}\lambda_u^{J} \nn \\
&\le \frac 12 \sum_{J\in\widetilde B_u^{\perp}}\widetilde\lambda_u^J +
\sum_{J\in B_u^{\perp}}\lambda_u^{J}.
\end{align}
Combining this with Eq.\ (\ref{above}) gives
\begin{align}
\sum_{J\in\widetilde B_u}\!\widetilde\lambda_u^J &\ge 1-\sqrt{1-F(\ket{\widetilde
\Phi_u},\ket{\Phi_u})} -\frac 12 \! \sum_{J\in \widetilde B_u^{\perp}}
\! \widetilde\lambda_u^J - \!\! \sum_{J\in B_u^{\perp}} \! \lambda_u^{J}
\nn \\ &\ge 1-2\sqrt{1-F(\ket{\widetilde\Phi_u},\ket{\Phi_u})}-2 \!
\sum_{J\in B_u^{\perp}} \! \lambda_u^{J}.
\end{align}
Determining the expectation value over $u$ gives
\begin{equation}
\left\langle\sum_{J\in \widetilde B_u}\widetilde\lambda_u^J\right\rangle \ge
1-4\sqrt\epsilon-2\epsilon.
\end{equation}

The dimension of the space used is $2^{n[S(\bar\rho)+\delta]}$, and the maximum
$\widetilde\lambda_u^J$ is no larger than $2^{-n(\bar S-2\delta)}$. Therefore it
is possible to remotely prepare the state $\ket{\widetilde\Phi_u}$ with classical
communication $n[S(\bar\rho)-\bar S+3\delta]+O(\log n)$ bits, and probability of
success at least $1-2^{-n[S(\bar\rho)-\bar S+3\delta]}$. Because
$\langle\sum_{J\in \widetilde B_u}\widetilde\lambda_u^J\rangle \ge 1-4
\sqrt\epsilon-2\epsilon$, the average fidelity for a success is at least
$1-4\sqrt{\epsilon}-2\epsilon$. The average fidelity including failures must be
at least $1-4\sqrt{\epsilon}-2\epsilon-2^{-n[S(\bar\rho)-\bar S+3\delta]}$.

Because the average fidelity between the states $\ket{\widetilde\Phi_u}$ and
$\ket{\Phi_u}$ is at least $1-4\epsilon$, it is easy to see from the triangle
inequality for fidelities that the states $\ket{\Phi_u}$ may be prepared with
average fidelity arbitrarily close to 1. Therefore we see that the states
$\ket{\Phi_u}$ may be prepared with fidelity arbitrarily close to 1 and
communication per prepared state arbitrarily close to the Holevo information
$S(\bar\rho)-\bar S$.

The situation that we have considered, where classical communication is the
resource of interest and entanglement is a free resource, is analogous to that
of entanglement concentration and dilution \cite{condil}, where entanglement is
the resource under consideration and classical communication is a free resource.

The results of Refs.\ \cite{kholevo,schuwest,holevo} show that it is possible to
perform classical communication equal to the Holevo information (analogous to
entanglement concentration). Here we have shown that it is possible to remotely
prepare ensembles of mixed states using communication equal to the Holevo
information (analogous to entanglement dilution).

One consequence of our proof is that it is possible to convert between multiple
copies of different ensembles with the same total Holevo information
\cite{ben}, in an analogous way as it is possible to convert between different
pure entangled states via entanglement concentration and dilution. These
analogies are summarized in Table \ref{table}.

An important application of our optimal remote state preparation scheme is given
in Ref.\ \cite{ben}. Ref.\ \cite{ben} shows that, provided it is possible to
efficiently prepare ensembles, the classical communication capacity of a unitary
operation in a single direction is equal to the maximum by which the operation
may increase the Holevo information of an ensemble. This result is important
because it makes the evaluation of the communication capacity of an operation
tractable.

It is interesting to speculate whether the same is true for bidirectional
communication. In this case we would generalize to a bidirectional ensemble
$\{p_i,q_j,\ket{\Phi_{ij}}\}$, where $i$ is chosen by Alice and $j$ is chosen by
Bob. As discussed in Refs.\ \cite{berry1,berry2}, the same is true in the
bidirectional case if it is possible to create bidirectional ensembles using as
much communication as can be performed using these ensembles. This problem is a
topic for future research.

The authors thank Stephen Bartlett for suggestions on the manuscript.
This project has been supported by the Australian Research Council.

{\it Note}:
Bennett {\it et al.\ }\cite{ben} refer to a private communication from Peter
Shor claiming a proof similar to that shown here for optimal remote state
preparation.

\end{document}